\documentclass[aps,pre,showpacs,twocolumn]{revtex4-1}

\usepackage{graphicx}
\usepackage{dcolumn}
\usepackage{bm}
\usepackage[utf8]{inputenc}
\usepackage[T1]{fontenc}
\usepackage{mathptmx}
\usepackage{etoolbox}
\usepackage{CJK}

\usepackage{ulem}
\usepackage{amssymb}
\usepackage{epsfig}
\usepackage{subfigure}
\usepackage{mathrsfs}
\usepackage{verbatim}
\usepackage{rotating}
\usepackage{longtable}
\usepackage{amsmath}
\usepackage{array,color}

\begin{document}
\title{Amplifier scheme: driven by indirect-drive under $\sim$ 10 MJ laser toward inertial fusion energy}

\begin{CJK}{GB}{gbsn}
\author{Yongsheng Li(ÀîÓÀÉý)}
\author{Ke Lan (À¶¿É)}\email{lan\_ke@iapcm.ac.cn}
\author{Hui Cao (²Ü»Ô)}
\author{Yao-Hua Chen (³ÂÒ«èë)}
\affiliation{Institute of Applied Physics and Computational Mathematics, Beijing 100094, China}
\author{Xiaohui Zhao (ÕÔÏþêÍ)}
\author{Zhan Sui (ËåÕ¹)}
\affiliation{Shanghai Institute of Laser Plasma, China Academy of Engineering Physics, Shanghai 201800, China}

\begin{abstract}
Burn  efficiency $\Phi$ is a key for  commercial feasibility of fusion power station for inertial fusion energy, while $\Phi$ is usually  lower than  30$\%$  in the central ignition scheme of inertial confinement fusion (ICF).
A recent conceptual design for a  10 MJ laser driver [Z. Sui and K. Lan et al., {\it Matter Radiat. Extremes} {\bf 9}, 043002 (2024)] provides a new room for  target design to achieve a higher $\Phi$.
Here, we take the advantage of fuel density in reaction rate and propose a novel amplifier scheme for increasing $\Phi$   via two cascading explosions by ICF.
The amplifier scheme can be realized either by indirect-drive or by direct-drive.
Here, we give a 1D design for an indirect-driven amplifier capsule  containing 2.02 mg DT fuel under a 300 eV radiation generated by a 10 MJ and 1785 TW laser inside an octahedral spherical hohlraum.
At stagnation, it forms an extremely dense shell surrounding central hot fuel,  with a  density ratio of shell to central  $>$ 20.
About 53 ps after stagnation, benefiting from the extremely high density of the shell and the deposition of $\alpha$ particles generated in the central hot fuel, the primary explosion happens in the shell.
Then, the  primary explosion in shell drives the central fuel to converge spherically towards the center.
At about 18 ps after the primary explosion, the central fuel converges at  center with
1100 g/cm$^3$,  770 keV and 320 Tbar, leading to the secondary explosion inside this extremely hot and dense fireball.
As a result, the  amplifier capsule has $\Phi$ = 48$\%$ and G = 33 at convergence ratio $C_r$ = 24. This novel scheme can achieve a relatively high burn efficiency at a relatively low $C_r$, which can greatly relax the stringent  requirements of high gain fusion on  hot spot ignition conditions and engineering issues.
\end{abstract}

\pacs{52.57.Fg, 52.35.Py, 52.38.Mf}
\maketitle
\end{CJK}


Fusion has the potential to provide a reliable, limitless, safe, and clean energy source \cite{MTV, IFEneedsreport},
and the successful achievement of ignition for indirect-drive inertial confinement fusion (ICF)
\cite{Abu-Shawareb2024PRL, Hurricane2024PRL, Rubery2024PRL, Pak2024PRE} at the National Ignition Facility (NIF) \cite{Lindl1995POP, Maclaren2014PRL, Betti2016NP, Campbell2017MRE} makes the inertial fusion energy (IFE) a highly promising approach.
However,   the target gain G required by IFE, defined as the ratio of fusion energy output to  laser energy on target, is estimated to be   30 - 100  to achieve attractive economic performance \cite{MTV}, much higher than the currently recorded highest  G $\sim$ 2.4  on the NIF\cite{5.2MJ}.
Driven by laser energy $E_d$, G of fuel with mass $m_{fuel}$ can be expressed as:
\begin{eqnarray}\label{Eq1}
G = \Phi \times Q \times m_{fuel}/E_d.
\end{eqnarray}
Here, $Q$ is the released energy by a nuclear reaction per unit mass and $\Phi$ is   burn  efficiency.  Usually, $\Phi$  is smaller than $30\%$ in the central ignition scheme of ICF. Obviously, $\Phi$ is a key to increase G at given $m_{fuel}$ and $E_d$.

An equal molar mixture of deuterium and tritium (indicated as symbols D and T, respectively)  considered in the majority of present fusion research  has the most important reaction

D + T $\rightarrow$ $\alpha$  (3.52 MeV) + n (14.1 MeV)
\\
for fusion research due to their largest fusion cross-section.
As known, DT reactivity is $\propto$  $T_i^2$ in the range of 8 - 25 keV, here $T_i$ is ion temperature.
Thus, the fusion is ignited in the hot central fuel in the central ignition scheme \cite{MTV}.
On the other hand, we have the volumetric reaction rate of DT   as
\begin{eqnarray}\label{Eq1}
R_{\texttt{DT}} \propto  \frac{\rho^2}{\bar{m}^2} <\sigma v>,
\end{eqnarray}
where $\rho$ is mass density, $\bar{m}$ is the average nuclear mass, $<\sigma v>$ is averaged reactivity.
It shows a very important feature for fusion energy research: $R_{\texttt{DT}} \propto$ $\rho^2$, or the reaction rate per unit mass is proportional to $\rho$, indicating the role of $\rho$ of the fuel in achieving efficient release of fusion energy and high burn efficiency \cite{MTV}.

\begin{figure}[htbp]
 \centering
\includegraphics[width=0.49\textwidth]{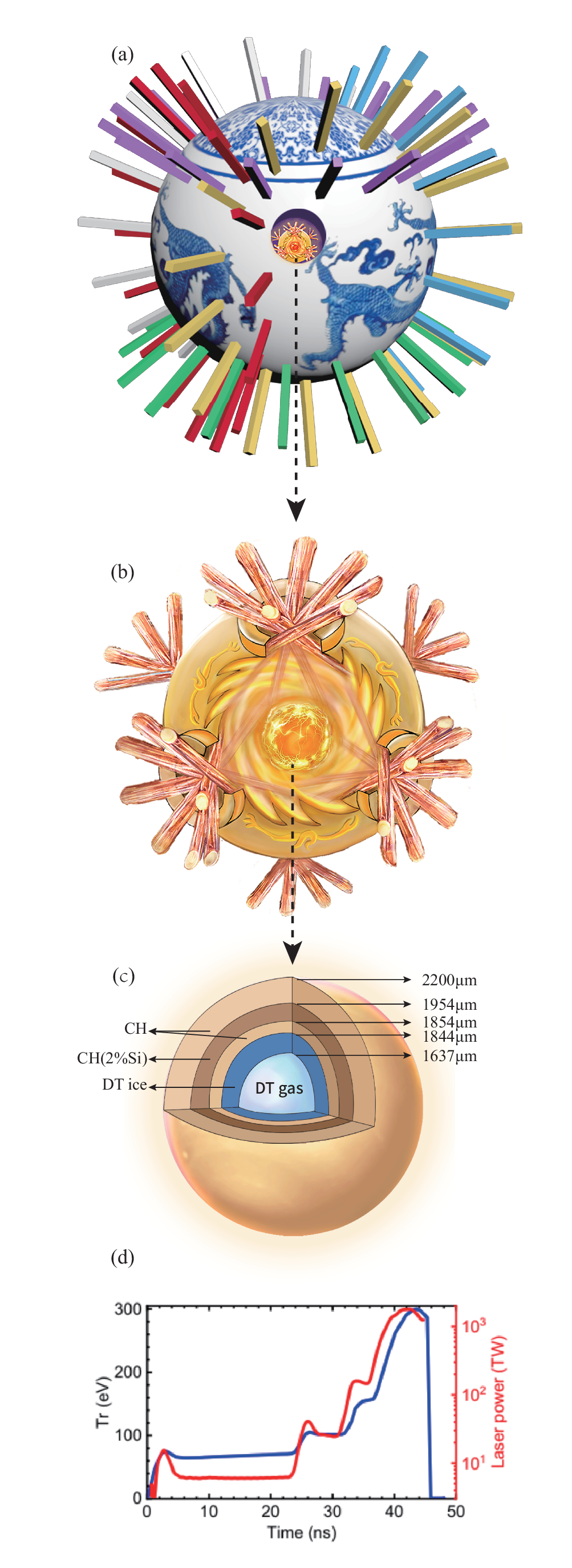}
\caption{(Color online) Artistic representations of the target chamber in octahedral configuration (a),  octahedral spherical hohlraum  (b), spherical CH capsule that contains DT fuel (c), and laser pulse (red) and radiation drive (blue)(d).}
\label{Fig:1}
\end{figure}

A recent conceptual design\cite{10MJ} for a  10 MJ laser driver  provides a new room for novel target designs for IFE.
In this paper, we will take  above advantage of $\rho$ and propose a novel  amplifier scheme to increase the  burn efficiency  via two cascading explosions  at a relatively low convergence ratio under 10 MJ laser.
In contrast to the central ignition scheme with only one explosion in the central hot fuel, our novel scheme requires an extremely dense shell to be formed at stagnation, and in return, it has two explosions, with the primary one happening in the extremely dense shell and the secondary one happening in an extremely hot and dense central dense fuel generated by the primary one.
This amplifier scheme can be realized either by direct-drive or by indirect-drive.
A direct-driven amplifier design is given in our separate paper \cite{Lan2024POP}.
Here, we  present an indirect-driven amplifier design   with a spherical CH capsule inside an octahedral spherical hohlraum \cite{Lan2013, Lan2014a, Lan2014b, Lan2014c,  Wang2019, Craxton2020DPP, Craxton2021DPP, WangLLERev2021, WangWY2021, Marangola2021, Davies2022} driven by  10 MJ laser, and discuss and illustrate the principles of the novel  amplifier scheme by simulation results.
In this design, we use the most economical CH as  ablator and adopt the low entropy target design for the purpose of efficient compression and high burn efficiency of fuel, since the benefit-cost ratio should be considered to aim at energy production.

To design the capsule, we use  a 1D capsule-only multigroup radiation hydrodynamic code RDMG \cite{Feng1999, Lan2017POP, Qiao2019PPCF, Qiao2021PRL} to simulate the implosion dynamics.
We consider a radiation drive of four steps, with a 6 ns main pulse peaking at 300 eV.
As a result, the spherical CH capsule contains three layers of CH ablator, including undoped CH, 2$\%$ Si doped CH and undoped CH.
The capsule outer radius is $2200$ $\mu m$, ablator thickness is $356$ $\mu m$, and DT-ice layer thickness is $207$ $\mu m$.
The initial density is 1.069 g/cm$^3$ for CH, 1.147  g/cm$^3$ for 2$\%$ Si doped CH,   0.3 mg/cm$^3$ for DT gas, and 0.255 g/cm$^3$ for DT ice. The DT mass is 2.02 mg, and the total ablator mass is about 19.75 mg.
Hereafter, we call this capsule as the amplifier capsule.
To convert 3D lasers into a 1D spherical radiation   without symmetry tuning,
we consider an octahedral spherical hohlraum \cite{ Lan2022MRE, Chen2022MRE} with
a hohlraum-to-capsule radius ratio of  4 and six 2000-$\mu$m-radius laser entrance holes.
We use a  sandwich hohlraum wall \cite{Sandwich2010LPB}, which has been successfully applied in the NIF ignition experiments \cite{Haan2011, Callahan2012POP, Lindl2012LLNL, Kline2013POP, Doppner2015PRL, Kritcher2022PRE, Kritcher2024PRE}.
For simplicity, here we use an initial design method \cite{Lan2010LPB, Lan2012LPB, Lan2016MRE} to give  the temporal profile of laser pulse.
Laser absorption efficiency is taken as 95$\%$, by assuming we have a low-LPI  at the next generation laser system \cite{incoherent2023MRE}.
As a result,  a drive laser with 10 MJ  energy  and 1785 TW  peak power is required.
Shown in Fig. \ref{Fig:1} is the artistic representations of  target chamber, hohlraum, capsule and drive profiles.
In this paper, we mainly focus on the  implosion dynamics. The details of hohlraum design from a  two-dimensional (2D) multigroup radiation hydrodynamic code LARED-INTEGRATION \cite{Lan2005LPB, Yong2013, Lan2021PRL} will be presented in our forthcoming publications.

\begin{figure}[htbp]
 \centering
	\includegraphics[width=0.45\textwidth]{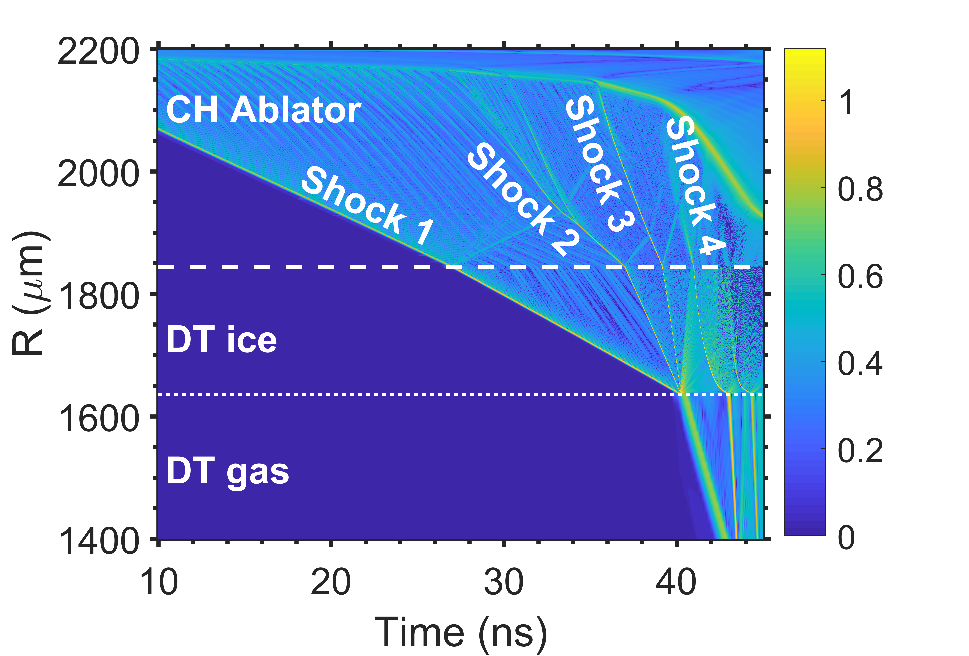}\hspace{0.5cm}
	\caption{\label{fig:shockTiming}(Color online) Plots of the logarithmic radial derivative of hydrodynamic pressure in Lagrangian coordinate vs time space. The green dashed line is the interface between CH ablator and DT ice, and the red dashed line is the interface of the DT ice/gas.}
\label{Fig:2}
\end{figure}

Presented in  Fig.\ref{Fig:2}  is   shock trajectories within the amplifier capsule, which are set off successively according to Munro criterion \cite{Munro2001}. As a result,   main fuel is compressed low adiabatically with a main fuel adiabat   $\alpha  \sim 1.46$.
As shown, each step of the radiation drive launches an inward shock, with   first three shocks merging at the interface of DT ice/gas, and the fourth one catching up with the former shocks within   DT gas, forming a much stronger shock. As the strong shock propagates within   DT gas, it will distribute thermal energy among ions and electrons according to their masses \cite{Fan2017MRE}.

\begin{table}\label{tab:1dips}
	\caption{Comparisons of   1D implosion parameters of the  amplifier capsule  and the NIC-Rev5 CH target.}\label{c}
	\begin{ruledtabular}
		\begin{tabular}{lrr}
		
			  Paramters                  & Amplifier & NIC-Rev5\\
			  \hline
			Drive laser Energy(MJ)/Power(TW)            &10.0/1785    & 1.35/415 \\
            Peak radiation (eV)             &300    & 300  \\
            Total length of radiation pulse (ns)             &45    & 21  \\
            Duration of main pulse  (ns)             &6    & 3.2  \\
			Capsule Outer Radius($\mu$m)  & 2200   & 1108 \\
			Ablator Mass (mg)            & 19.75  & 6.1     \\
			Fuel Mass (mg)               & 2.02   & 0.17 \\
			Absorbed capsule energy (MJ) & 0.98   & 0.16  \\
            Main Fuel Adiabat            & 1.46   & 1.40  \\
            Peak implosion Velocity (km/s)    & 300   & 370  \\
			Ablator Mass Remaining  (AMR)     & 14.5\%   & 9.4\%   \\
            Convergence ratio $C_R$           & 24  & 33   \\
			$(\rho R)_{H}$ at $t_{stag}$ (g/cm$^2$) & 0.51   & 0.47    \\
            Averaged $(\rho R)_{fuel}$ (g/cm$^2$) & 2.30   & 1.19    \\
            Burn  efficiency                   & 48\%     & 30\% \\
            Yield (MJ)                   &327   & 17.4 \\
			Target gain                  &32.7   & 13 \\
		\end{tabular}
	\end{ruledtabular}
\end{table}

In order to compare the main differences  between the amplifier scheme and the central ignition scheme, we also simulate the NIC-Rev5 CH capsule \cite{Haan2011} in  central ignition scheme.
The NIC-Rev5 CH capsule  contains 0.17 mg fuel and has a similar main fuel adiabat of 1.4 from our simulations.
In the Table I, we compare the 1D implosion performance parameters between the two capsules.
Drive laser energy of 1.35 MJ and power of 415 TW are simply taken from  Ref. \citenum{Haan2011}.
As shown, peak implosion velocity $v_{imp}$ of the amplifier capsule is  300 km/s, obviously slower than 370 km/s   of the NIC-Rev5  capsule.
AMR at $v_{imp}$  is $14.5\%$   and $9.4\%$ for the  amplifier  and   NIC-Rev5  capsules, respectively.
A higher AMR   means a thicker ablator, which can lead to a more hydro-stable fuel/ablator interface for the amplifier capsule.
$C_R$ is 24 for the amplifier capsule, obviously lower than $C_R$ = 33 of the  NIC-Rev5  capsule.
At stagnation time $t_{stag}$, areal density of hot spot $(\rho R)_{H}$ is similar for both capsules.
However, under the amplifier scheme, the averaged fuel areal density  $(\rho R)_{fuel}$ of the  amplifier capsule reaches 2.3 g/cm$^2$,   about twice as that of the NIC-Rev5 CH capsule,  which  guarantees  a higher burn efficiency $\Phi$ of the amplifier capsule despite its lower implosion velocity.
It seems not a fair comparison because  the amplifier capsule uses 7.4 times laser energy of the central ignition capsule, but note it is used for driving the 11.9 times fuel mass.
As a result, we have $\Phi$ = 48$\%$ with a fusion energy yield   Y$_{id}$ = 327 MJ and G = 32.7 for the amplifier capsule, and  $\Phi$ = 30$\%$, Y$_{id}$ = 17.4 MJ and G = 13  for the NIC-Rev5 capsule.

\begin{figure}[htbp]
	\centering
	\includegraphics[width=0.45\textwidth]{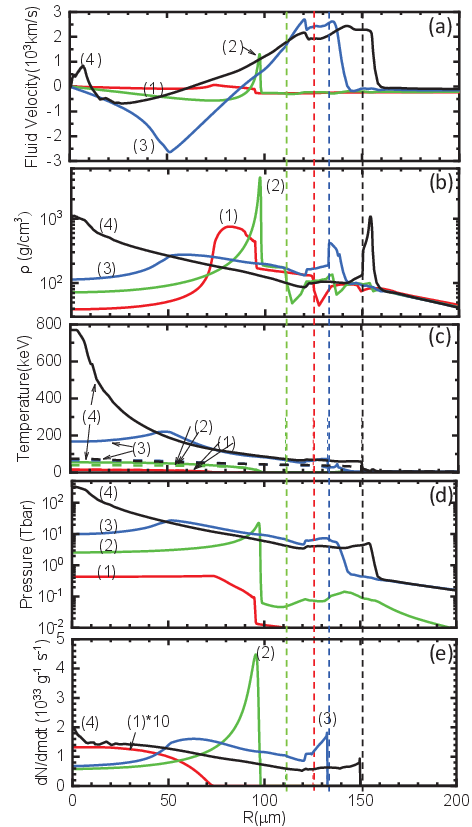}
\caption{\label{fig:Profiles}(Color online) Sequences of radial profiles of $v$ (a),
$\rho$ (b), $T_i$ (solid line) and $T_e$ (dashed line) (c), $P$ (d)
 and  $\frac{dN}{dmdt}$ (e)
at: (1) $t_{stag}$ (red), (2) $t_{pri}$ (green), (3) $t_{pri-sec}$ (blue) and (4) $t_{sec}$ (black).
The vertical thin dashed lines show the corresponding fuel/ablator interface, which continues coasting inward at a velocity of 260 km/s after $t_{stag}$ while  abruptly moves outward due to the primary explosion.
Note the steep changes of $\rho$ at the fuel/ablator interface at all the four times in (b).
At $t_{stag}$  and $t_{pri}$, $\rho$ dips in CH ablator at the interface, because CH ablator has a higher opacity than DT fuel and hence has a stronger absorption of radiation emitted by hot spot.
At $t_{pri-sec}$ (blue) and $t_{sec}$, $\rho$ rises abruptly in CH ablator at the interface, because it has a much lower temperature in CH ablator than in DT fuel. Note DT fuel is strongly heated by $\alpha$-particle deposition.
}
\label{Fig:3}
\end{figure}

Same as the central ignition scheme, the amplifier scheme includes implosion and stagnation, with fusion starting from the central hot spot and serving as a spark plug for ignition.
However, the fuel burn in the amplifier scheme is dominated by density and has following characteristics. First, an extremely compressed shell is required to be formed with  a very high density ratio of cold  shell and hot spot at stagnation when imploding material is stopped  and  comes to rest. Second, the extremely dense, cold and thick shell  completely stops the $\alpha$-particles generated in the central hot fuel and is rapidly heated up by $\alpha$-particle deposition, and when   meeting the ignition condition, the primary explosion happens in the middle of  shell. Third, the primary explosion violently splits the whole fuel into two parts,  pushing the outer part to expand while compressing the inner part to converge spherically, and the secondary explosion happens when the central fuel converges spherically at center.

We define three characteristics times for the amplifier scheme, including the stagnation time $t_{stag}$ when  kinetic energy of fuel in the shell attains its minimum,  the primary explosion  time $t_{pri}$ when $\frac{dN}{dmdt}$ reaches  peak in the extremely dense shell, and  the secondary explosion time $t_{sec}$ when $\frac{dN}{dmdt}$ reaches its peak at the fuel center. Here, $N$ is neutron number, $m$ is mass, $t$ is time,  $\frac{dN}{dmdt}$ is reaction rate of neutron per unit mass.
From  simulations, we have $t_{stag}$ =  47.400 ns,  $t_{pri}$ = 47.453 ns,  and $t_{sec}$ = 47.471 ns for the amplifier capsule, with  differences of 53 ps and 18 ps  between adjacent times.
In the following discussions, we also consider the case at $t_{pri-sec}$ = 47.464 ns, a selected time between  $t_{pri}$ and $t_{sec}$, to understand the plasma status between the primary and secondary explosions.
In Fig.\ref{Fig:3}, we present the radial profiles  of $v$, $\rho$, $T_i$, $T_e$, $P$  and $\frac{dN}{dmdt}$ of the amplifier capsule at the four times.
Here, $v$ is fluid velocity, $T_e$ is electron temperature, and $P$ is  pressure.

At $t_{stag}$, as shown in Fig.\ref{Fig:3} (a),  an extremely dense shell has been formed with  $\rho_{shell}$ $\sim$ 780 g/cm$^3$ and $\rho_{shell}$/$\rho_{center}$ $>$ 20, as shown in  Fig.\ref{Fig:3} (b);  $T_i$ and $T_e$ are  in  equilibrium, $\sim$ 14 keV, changing  little in whole hot spot, as shown in  Fig.\ref{Fig:3} (c); the whole hot spot area is isobaric with $P$ $\sim$ 0.42 Tbar,
and  $P$  drops rapidly as $\rho$ in the dense shell, as shown in  Fig.\ref{Fig:3} (d). Here, $\rho_{shell}$ denotes the peak density in  shell, roughly locating in the middle of shell,
and $\rho_{center}$ is the density at $R = 0$, the center of the spherical fuel.
As shown in  Fig.\ref{fig:Profiles}(e), $\frac{dN}{dmdt}$ is
flat in the central fuel, but decreases obviously in the inner boundary of  dense shell where $T_i$ decreases and $\rho$ increases rapidly.

At  $t_{stag}$, we define the hot spot boundary as the place where $\frac{dN}{dmdt}$ falls to 1$\%$ of its peak, and the shell ranges from the hot spot boundary to  the place of the shock front where $\rho$ in fuel jumps down.
According to this definition, the hot spot has a radius of 75.4 $\mu$m  and the shell has a width of 20  $\mu$m at  $t_{stag}$.
The hot spot is 0.198 mg in mass, only   10$\%$ of whole fuel mass.
In contrast, the shell is 1.148 mg in mass, about    57$\%$ of whole fuel mass.
Nevertheless, the internal energies of hot spot and shell are 120 kJ and 44.4 kJ, respectively.
It means that the internal energy per mass of hot spot is about 16 times that of the shell at this time.


At $t_{pri}$, benefiting from the extremely high density of shell and the deposition of $\alpha$ particles generated in the central hot fuel, the primary explosion happens in the shell when it meets the $\rho RT$  ignition condition \cite{MTV}. It is particularly interesting  that  the primary explosion picture of  amplifier scheme is quite different from  the central ignition scheme. In the latter,  explosion happens in the central hot fuel and whole fuel expands immediately after explosion.  In contrast,  the primary explosion of the amplifier scheme happens in the middle of the extremely dense cold shell and violently splits the whole fuel into two parts, as shown in Fig.\ref{Fig:3} (a), pushing the outer part to expand while  compressing the inner part to converge spherically to form an extremely dense and hot fireball.
At this time, $\rho_{shell}$  $\sim$ 4350 g/cm$^3$ with $\rho_{shell}$/$\rho_{center}$ $\sim$ 60, as shown in Fig.\ref{Fig:3} (b); $P_{shell}$ $\sim$ 22 Tbar with $P_{shell}$/$P_{center}$ $\sim$ 9,   as shown in Fig.\ref{Fig:3} (d).
Note that in the central ignition scheme, the shell pressure is never significantly higher than in central hot fuel and it cannot form intense combustion in the  fuel shell.
From Fig.\ref{Fig:3} (e), $\frac{dN}{dmdt}$ reaches its peak of  4.5 $\times$ $10^{33}$ s$^{-1}$g$^{-1}$  at $R$ = 95.7 $\mu$m in the middle of shell, around where $\rho$ peaks at 4400 g/cm$^3$ and $P$ peaks at 22 Tbar.
At this time, $\frac{dN}{dt}$ of whole fuel also reaches its peak of $10^{31}$ s$^{-1}$.
From Fig.\ref{Fig:3} (c), non-equilibrium between ion and electron \cite{Fan2017MRE} with $T_i/T_e$ = 1.4 can be clearly seen in the central fuel. Note that $T_i$ drops to 13 keV at $R$ = 95.7 $\mu$m where $\frac{dN}{dmdt}$ peaks. Obviously, the primary explosion is dominated by density.

\begin{figure}[htbp]
 \centering
	\includegraphics[width=0.45\textwidth]{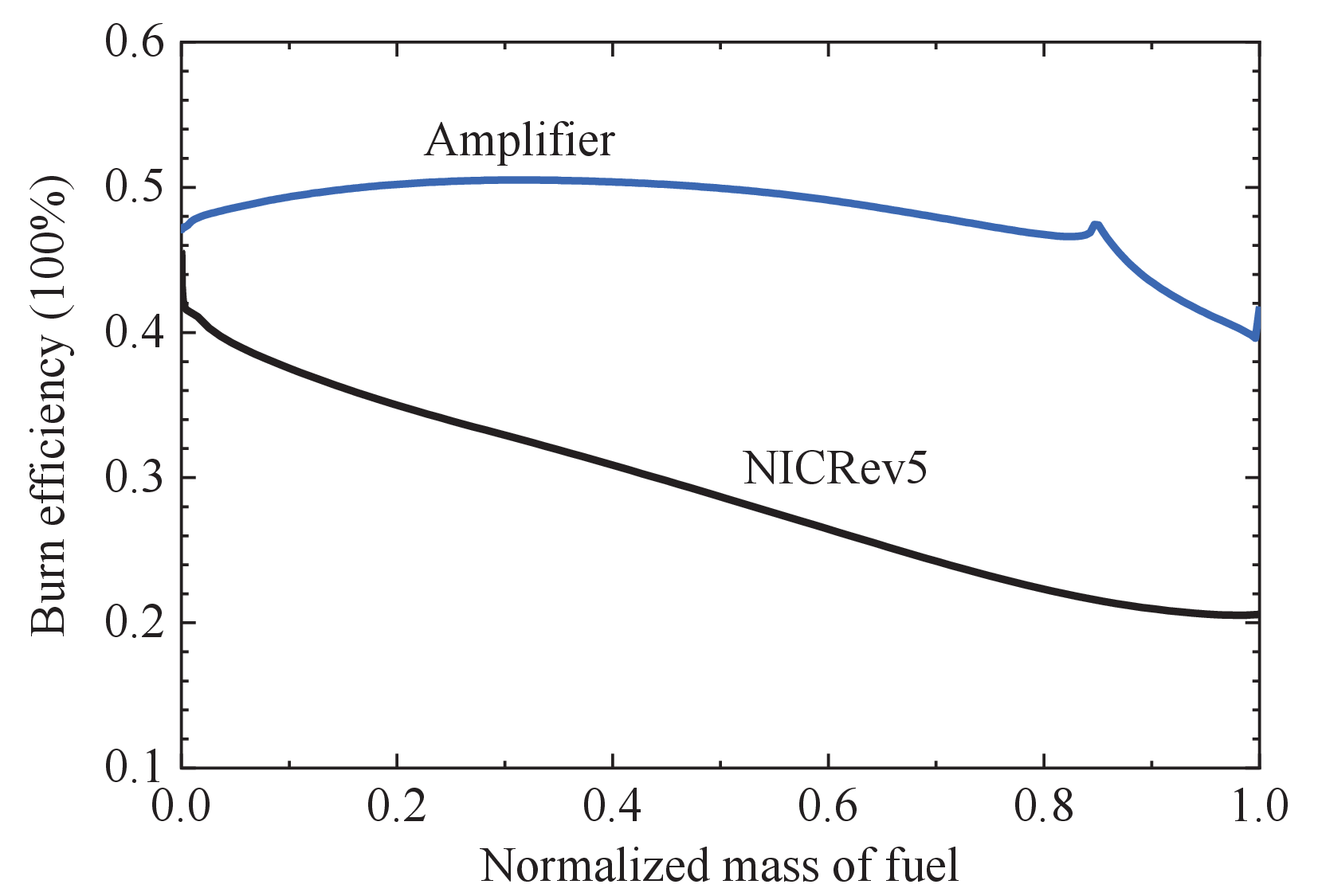}\hspace{0.5cm}
	\caption{\label{fig:shockTiming}(Color online) Variation of burn efficiency along radial direction in fuel. The horizontal axis is the normalized  mass within  radial position to the total fuel mass.}
\label{Fig:4}
\end{figure}

At $t_{pri-sec}$, under the huge fusion power released by the primary explosion, both  implosion of the inner part  and  explosion of the outer part of the dense shell becomes so strong that, as shown in Fig.\ref{Fig:3} (a), $|v|$  exceeds $\sim$ 2600 km/s, about 9 times  the implosion velocity  under the  300 eV radiation generated in hohlraum.
It leads to the violent  decrease/increase  of $\rho$ in the outer/inner part of fuel,
as  shown in Fig.\ref{Fig:3} (b).
Such as, compared with $t_{pri}$, $\rho$ at $R = 0$ increases from 72 to 114 g/cm$^3$, while $\rho$ at $R$ = 95.7 $\mu$m decreases from 4400 to 200 g/cm$^3$.

From Fig.\ref{Fig:3} (c), $T_i$  in the fireball increases   abruptly, which can be contributed by mechanical work via compression and $\alpha$-particle deposition produced in the primary explosion.
From Ref. \citenum{MTV}, we can estimate $W_{dep}$ and $W_m$,  respectively, with:
\begin{equation}
W_{dep} = 1.54 \times 10^{-31} \eta_{dep} n^2 T_i^2/\rho  ~~ \text{J/s/g}  \text{,}
\end{equation}
and
\begin{equation}
W_{m} =  \frac {3 P u} {\rho R} \text{,}
\end{equation}
where  $W_{dep}$ is the $\alpha$-particle deposition power per mass,
$W_m$ is the mechanical work power per mass, $\eta_{dep}$ is deposition factor of $\alpha$-particle, ion density $n$   in cm$^{-3}$, temperature $T$ in keV, $\rho$ in g/cm$^3$,  $P$ is pressure, $u$ is  velocity, and $\rho R$  is areal density.
At $t_{pri}$, our simulation gives the averaged $\rho R$ = 1.56 g/cm$^2$, $\rho$ = 268 g/cm$^3$, and $T_i$ = 45 keV for the fireball.
By using the expressions in Ref. \citenum{MTV}, we can  estimate that the range of $\alpha$-particle is  0.0056 cm and the deposition factor is 77$\%$ for this case.
Here,
we use the following expression of Ref. \citenum{Li2019POP} to calculate $\eta_{dep}$:
\begin{widetext}
\begin{equation}
\eta_{dep} = 1 - \frac{0.00593(\rho R)^{-1.174}T^{1.556}}{1+0.00385(\rho R)^{0.600}T^{1.316}+0.00547(\rho R)^{-1.180}T^{1.574}} \text{,}
\label{eta_t_fit}
\end{equation}
\end{widetext}
which considers  all modifications of the $\alpha$-particle stopping by both DT ions and  electrons with their Maxwellian average stopping weights, the relativity effect on electron distribution and the modified Coulomb logarithm of DT-$\alpha$ collisions and gives a smaller deposition factor than that in
Ref. \citenum{MTV}.
Then, we can have $\eta_{dep}$ = $57\%$ from Eq. (5) and $W_{dep}$ = $2.7 \times 10^{21}$ J/s/g from Eq. (3), which approximately agrees with $1.6 \times 10^{21}$ J/s/g from our simulation.
For $W_{m}$, we take $P$ as the pressure difference between the fireball boundary and center, $u$ the implosion velocity of fireball boundary, and $R$ the fireball radius. So we have  $P$ $\sim$ 20 Tbar, $u$ $\sim$ 5.5 $\times 10^7$ cm/s from our simulations.
Then, we have  $W_{m}$ $\sim$ $2.3 \times 10^{18}$ J/s/g from Eq. (4), approximately  agreeing with $2.14 \times 10^{18}$ J/s/g from our simulation.
Hence, $W_{dep} \gg W_{m}$, indicating the abrupt increase of $T_i$  in the fireball is mainly due to  the very strong energy deposition of $\alpha$ particles produced in the primary explosion.
Considering the specific heat $C_{vi}$ = 5.79 $\times 10^7$ J/g/keV for  DT  and assuming that half of  the deposition energy at boundary is given to the fireball, the increase of $T_i$ within 11 ps from $t_{pri}$ to $t_{pri-sec}$ is about 260 keV, approximately agreeing with the results in Fig.\ref{Fig:3} (c).

Note at   $t_{pri-sec}$, we have $T_i$ = 170 keV while $T_e$ = 66 keV at $R = 0$, indicating a very strong non-equilibrium between ions and electrons at this time.
At   $t_{pri-sec}$, it is interesting to note from Figs. \ref{Fig:3} (a)-(e) that implosion velocity  peaks at 2640 km/s at R $\sim$ 50 $\mu$m. Simultaneously, at this place, $\rho$ also peaks at 275 g/cm$^3$, $T_i$ peaks at 207 keV,  $P$  peaks at 27 Tbar, $T_i/T_e$ reaches  3.9, and $\frac{dN}{dmdt}$  reaches  1.6 $\times$ $10^{33}$ s$^{-1}$g$^{-1}$.
Especially, $\frac{dN}{dmdt}$ at the fuel/ablator interface reaches   1.7 $\times$ $10^{33}$ s$^{-1}$g$^{-1}$, the highest in the whole fuel,  indicating that whole fuel is burnt at this time.

At $t_{sec}$, the primary explosion generated extremely hot and dense fireball spherically converges at fuel center and the secondary explosion happens.
Around this time,  the fuel at center starts to expand,  as shown in Fig.\ref{Fig:3} (a).
From Figs.\ref{Fig:3} (b), (c) and (d),  all of $\rho$, $T_i$, $P$ and  $\frac{dN}{dmdt}$  reach their peaks  of 1100 g/cm$^3$,  770 keV, 320 Tbar, and $2 \times 10^{33}$ s$^{-1}$ g$^{-1}$ at R = 0,  respectively. It means that the secondary explosion benefits from both density and temperature.
At this time,   $\frac{dN}{dt}$ of whole fuel  reaches 1.6 $\times$ $10^{30}$ s$^{-1}$ and
$\frac{dN}{dmdt}$ is  9.3 $\times$ $10^{32}$ s$^{-1}$g$^{-1}$ at the fuel/ablator interface.


Presented in Fig.\ref{Fig:4} is a comparison of variation of $\Phi$ along radial direction in fuel between the amplifier capsule and NIC-Rev5 CH capsule.  As shown, $\Phi$ changes small and is within 40$\%$ and 50$\%$ in the whole fuel of the amplifier capsule, while it drops obviously from   center to boundary from 40$\%$ to 20$\%$ for NIC-Rev5 CH capsule.
We also compare the yield released before and after bang time when $dN/dt$ of whole fuel reaches its peak of the two kinds of capsules.
As a result, the yield released by the amplifier capsule  after bang time is 4.2 times that before, while it is 1.7 times for the  NIC-Rev5 CH capsule. From our simulation,  $t_{pri}$ is 2 ps earlier than bang time of the amplifier capsule,  and its yield released after $t_{pri}$ is 11 times that before.
It demonstrates that the amplifier capsule can release remarkable additional yield in burn stage after ignition and has a remarkably higher $\Phi$  via two cascading explosions than the central ignition capsule.


\begin{figure}[htbp]
	\centering
	\includegraphics[width=0.45\textwidth]{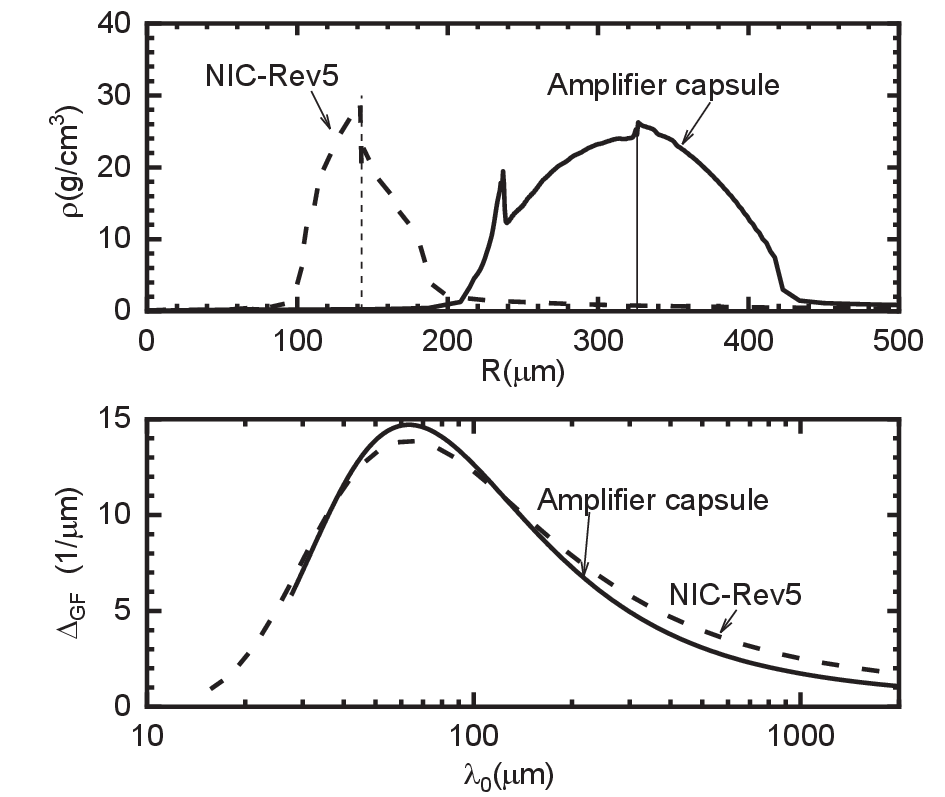}
	\caption{\label{fig:denProfile}(Color online) Radial profile  of $\rho$ in shell  (a)  and normalized RTI growth factor vs  disturbance wavelength at the initial surface of capsule (b) at $t_{imp}$  for the amplifier capsule (solid line) and the NIC-Rev5 capsule (dashed line), respectively.
Vertical thin  lines   in (a) mark the material interface between DT fuel and CH ablator of the two capsules.}
\label{Fig:5}
\end{figure}

Here, we simply discuss the hydrodynamic instabilities of the amplifier capsule.
As claimed above, we take a higher AMR in our design in order to have a more hydro-stable fuel/ablator interface and reduce mixing \cite{Goncharov1999PRL, Do2022PRL, Bachmann2022PRL}.
We optimize the design by increasing cautiously the thicknesses of ablator and doped layer, at the cost of reducing implosion velocity, to mitigate the hard X-ray preheat in order to increase the ablator density adjacent to the main fuel.
As a result, the density of main fuel is kept lower than the ablator until to $t_{imp}$, the time of the maximum implosion velocity before $t_{stag}$, as shown in Fig.\ref{Fig:5} (a).
It indicates that our design can keep the Atwood number being negative at the interface throughout the acceleration and ensure the stability of material interface.
The results of NIC-Rev5 capsule is also presented for comparison.

In addition, the ablation front linear growth factor (GF) of Rayleigh-Taylor hydro-instability (RTI) at the ablation surface  can   be obtained by using a simple linear analysis \cite{Lindl2004POP}.
We normalize GF to  the ablation layer thickness at $t_{imp}$ and denote it as $\Delta_{GF}$.
We present  $\Delta_{GF}$ in Fig.\ref{Fig:5} (b),  and it shows little difference  between the two capsules. As shown, the initial wavelength of the disturbance   grows most rapidly at the ablation surface $\sim$ 70 $\mu$m for both capsules.
From Ref. \citenum{Haan2011}, the corresponding mode is 120 and the surface disturbance amplitude of this wavelength  is  $\sim$ 1 nm for the NIC-Rev5 capsule with such an ablation surface. It indicates, even though it grows linearly until to that the shell reaches its maximum implosion velocity before $t_{stag}$, the amplitude is still much smaller than the ablation layer thickness and can be neglected.

Note it spends very short time of 18 ps from the primary explosion to secondary explosion, which is reasonable under a drive of primary explosion. Thus, it can be expected that degradation due to hydro instabilities will not seriously affect the performance of the second explosion.
Nevertheless, the requirement for a high density ratio of the cold shell to the hot spot in the amplifier
capsule may be challenging and lead to hydrodynamic unstable.
We will investigate the hydro instabilities of the amplifier capsule by considering   X-ray drive asymmetry,  supporting membrane,   fill tube,  local defects of the shell by 2D or 3D simulations in our future work.

In summary,  we have proposed a novel amplifier scheme for increasing burn efficiency via two cascading explosions by inertial confinement fusion and  presented an indirect-drive amplifier design with a spherical CH capsule inside an octahedral spherical hohlraum driven by  10 MJ laser.
Our simulation results on the  NIC-Rev5 CH capsule in central ignition scheme is also presented for comparison.
As a result, the  amplifier capsule has $\Phi$ = 48$\%$ and G = 33 at convergence ratio $C_r$ = 24, while it is $\Phi$ = 30$\%$ and G = 13 at $C_r$ = 33 for  the  NIC-Rev5 CH capsule.
It is worth mentioning that our amplifier scheme is very   different from the shock ignition scheme \cite{Betti2007PRL} which needs an ignitor shock to heat its central hot spot to ignite the assembled fuel.
In contrast, the amplifier scheme with two cascading explosions  can be realized fully under inertial confinement, with no need of any ignitor shock.
The detail differences between the amplifier scheme and the shock ignition scheme is presented in Ref. \citenum{Lan2024POP}.
The amplifier scheme  can happen at a relatively low convergence ratio, so it can    relax the stringent requirements on $\rho R T$ hot spot condition, drive asymmetry, laser-plasma instabilities, and hydrodynamic instabilities usually required by the central ignition scheme for a high gain fusion.
In the future, we will do the parameter scan for giving trigger criterions of the amplifier scheme and  optimize the amplifier design for a higher burn efficiency under a lower laser energy.

{\bf ACKNOWLEDGMENTS}
K. L.  appreciates Professor Vladimir Tikhonchuk of the ELI-Beamlines for beneficial discussions on our novel scheme and appreciate S. Atzeni and J. Meyer-ter-Vehn for their very nice book, Ref. \citenum{MTV} in helping to understand and describe the novel phenomena.
This work is supported by the National Natural Science Foundation of China (Grant No. 12035002).

\end{document}